\documentstyle[pra,aps]{revtex}
\begin{document}
\draft
\title{
SHORT-RANGE CORRELATIONS AND THE ONE-BODY DENSITY MATRIX
IN FINITE NUCLEI }
\author{A. Polls}
\address{Departament d'Estructura i Constituents de la Mat\`eria\\
Universitat de Barcelona\\
Diagonal 647, E-08028 Barcelona, Spain}
\author{H. M\"uther}
\address{
Institut f\"ur Theoretische Physik, Universit\"at T\"ubingen,\\
Auf der Morgenstelle 14, D-72076 T\"ubingen, Germany  }
\author{W.H. Dickhoff}
\address{
Department of Physics, Washington University,\\
St. Louis, MO 63130, USA}
\maketitle
\begin{abstract}
The effects of short-range correlations derived from a realistic
meson-exchange potential on the single-particle density matrix in
finite nuclei are investigated by analyzing the one-body density in
terms of the natural orbits. Basic features of these natural orbits and
their spectral distributions are
discussed. For many observables it seems to be
sufficient to approximate the one-body density matrix in terms of those
natural orbits, which exhibit the largest occupation probabilities. For the
investigation of the high-momentum components in the single-particle
density, however, it is important to take into account natural
orbits with small occupation probabilities, originating from the
single-particle Green function at large negative energies.
\end{abstract}

\section{Introduction}
It is generally accepted that atomic nuclei are many-body systems in which
correlations beyond the mean-field or Hartree-Fock picture play a
significant role. Therefore it has always been a point of experimental and
theoretical studies to find observables, which reflect these correlations
in a unambiguous way.
In this sense,
the simultaneous study of density and momentum distributions in
nuclei seems to be a sensitive test of nuclear models and may provide
information on the correlations between the interacting nucleons\cite{jami}.
Density and
momentum distributions are one-body properties and therefore can be
calculated from the single-particle spectral function which contains all
the information on the single-particle structure of nuclei. Recent
proton knock-out experiments employing the $(e,e'p)$ reaction have shown that
correlations between nucleons are responsible for a reduction of the
spectroscopic strength of valence hole states with respect to the
uncorrelated mean-field values and for the fragmentation of the strength
over a large excitation energy domain of the remaining
nucleus\cite{lapi,bobel,blom}.

The modern $(e,e'p)$ performed e.g. with the MAMI facility\cite{blom}
in Mainz or at NIKHEF\cite{bobel} in Amsterdam do not only provide
information on the global spectroscopic strength but also yield results on
the high-momentum components of the spectral function.
The effects of correlations on the momentum and energy distribution
of the single
particle strength, in particular for these high momentum values, is
presently a subject of discussion\cite{morita,ciofi,benpa,piep,mawa,mahau}.
Most of the investigations have been performed for light
nuclei\cite{morita,ciofi,benpa} or infinite nuclear
matter\cite{ramos,benff,ciofb,vonde,gearh,baldo,kohl}. Results
for heavier nuclei are typically derived from investigations of nuclear
matter assuming a local density approximation \cite{string,sffb,bffs,neck}.

A question that has attracted a
lot of attention concerns the possibility of describing a given momentum
distribution by means of an independent particle  model. In the IPM the
nucleus is considered to be a system of nucleons moving without residual
interaction in a mean-field or single-particle potential. In this case
the nuclear wave function is represented by a single Slater determinant.
One way to clarify this point is by studying the natural orbits
($no$)\cite{mahau,loew,anto,stoits,waroq}.
These wave functions and their associated eigenvalues ($n_{\alpha}$)
(natural occupation numbers) are obtained by diagonalizing the one-body
density matrix for the ground state of the A particle system. Therefore the
one-body density matrix of the correlated system can be expressed in terms
of these occupation numbers $n_{\alpha}$ and the corresponding natural
orbits $\phi^{no}_{\alpha}$ by
\begin{equation}
\rho(r,r') = \sum_{\alpha} n_{\alpha} {\phi^{no}_{\alpha}}^*(r)
\phi^{no}_{\alpha}(r')
\label{eq:natorb}
\end{equation}
As the one
body density matrix is diagonal in the $no$ representation, this basis
appears to contain the most suitable single-particle wave functions for the
calculation of the expectation value of a general one-body operator.
In particular, if the total wave function of the system is a unique
Slater determinant build with single-particle wave functions then the
natural orbits coincide with those wave functions and the natural
occupations are equal to unity. The deviations of this ideal situation
indicate the impossibility to obtain the one-body density matrix from
a single Slater determinant. One can say that the effect
of correlations in the calculation of one-body observables are smallest
when using the $no$ basis; for example, the depletion of the Fermi sea is
smallest in this basis\cite{mahau}.

Performing a microscopic calculation of the one-body Green function for
$^{16}$O we have recently demonstrated\cite{pap1,pap2} that the
nucleon-nucleon correlations induced by the short-range and the
tensor components of a realistic interaction\cite{rupr} yield an
enhancement of the momentum distribution at high momenta as compared to
the Hartree-Fock description. This enhancement, however, is only obtained
after integrating over all excitation energies of the $A-1$ particle
system. On the other hand, a momentum distribution very similar to the
one derived in the Hartree-Fock approximation is observed if one restricts
the energy integration to the states of lowest energy for each set of
orbital angular momentum $l$ and $j$. Does this imply that correlation
effects are visible in $(e,e'p)$ experiments to the
low-lying states in $A-1$ nucleus only in terms of the spectroscopic
factors, which are reduced as compared to the independent particle
model? In order to discuss this question, we
will analyze the results of the single-particle Green function obtained
with the method described in \cite{pap2} in terms of natural orbits and
their occupation numbers.
This analysis will also allow us to study the effects of correlations on
the local density distribution $\rho(r)$ and other one-body observables in
$^{16}$O.

After this introduction, the techniques to evaluate the single-particle
Green function, the density matrix and the natural orbits are
briefly described in the next section. The discussion of our numerical
results are
presented in section 3 and the main conclusions of the work
are summarized in the final section 4.

\section{Calculation of the Single Particle Density Matrix}

The aim of our investigation is the analysis of the single-particle
density matrix $\rho_{mn}$, which is defined in a basis of appropriate
single-particle states $m,n$ by
\begin{equation}
\rho_{mn} = <\Psi_0^A\vert a^\dagger_m a_n\vert \Psi_0^A>  \,
\end{equation}
where $\Psi_0^A$ denotes the correlated ground state of the nucleus and
$a^\dagger_m$ ($a_n$) define the single-particle creation (annihilation)
operators for the basis considered. For a nucleus like $^{16}$O with a ground
state of angular momentum $J=0$ the density matrix can be separated into
sub-matrices of given orbital angular momentum $l$, total angular momentum $j$
and an isospin quantum number. In our discussion we will ignore the effects
of the Coulomb interaction. This means that the results obtained in nuclei
with $N$ = $Z$ are identical for protons and neutrons and therefore we will
drop the quantum number referring to the isospin. Using e.g. the momentum
representation and performing a partial wave decomposition,
the density matrix can be written as $\rho_{lj}(k_1,k_2)$, with
$lj$ referring to the ``good'' quantum numbers of angular momentum and $k_1$,
$k_2$ define the absolute values of the momenta for the single-particle
states $m,n$ above.

Within the Green function approach\cite{wimref} this density matrix can
be evaluated from the imaginary part of the single-particle Green function
by integrating
\begin{equation}
\rho_{lj}(k_1,k_2) = \int_{-\infty}^{\epsilon_F} dE\,
\frac{1}{\pi} \mbox{Im} \left( g_{lj}(k_1,k_2;E) \right) ,
\end{equation}
where the energy variable $E$ corresponds to the energy difference between
the ground state of the $A$ particle system and the various energies of all
the states in the $A-1$ system
(negative $E$ with large absolute value correspond
to high excitation energies in the remaining system) and $\epsilon_F$ is the
Fermi energy.

The single-particle Green function $g_{lj}$ or propagator
is obtained from a solution of the Dyson equation
\begin{equation}
  g_{lj}(k_1,k_2;E) = g^{(0)}_{lj}(k_1,k_2;E)
+ \int dk_3\int dk_4 g^{(0)}_{lj}(k_1,k_3;E) \Delta\Sigma_{lj}(k_3,k_4;E)
 g_{lj}(k_4,k_2;E) ,
\label{eq:dyson}
\end{equation}
where $g^{(0)}$ refers to a Hartree-Fock propagator and $\Delta\Sigma_{lj}$
represents contributions to the real and imaginary part of the irreducible
self-energy, which go beyond the Hartree-Fock approximation of the nucleon
self-energy used to derive $g^{(0)}$. The definition and evaluation of the
Hartree-Fock contribution as well as the calculation of $\Delta\Sigma$ will
be discussed below.

The calculation of the self-energy is performed in terms of a $G$-matrix
which is obtained as a solution of the Bethe-Goldstone equation for
nuclear matter choosing for the bare NN interaction the One-Boson-Exchange
potential $B$ defined by Machleidt (\cite{rupr}, Tab.A.2). The Bethe-Goldstone
equation has been solved for nuclear matter with a
Fermi momentum $k_F = 1.4 fm^{-1}$. This roughly corresponds
to the saturation density of nuclear matter. The starting energy
has been chosen to be -10 MeV. The choices for the density
of nuclear matter and the starting energy are rather arbitrary. It turns
out, however, that the calculation of the Hartree-Fock term is not very
sensitive to this choice\cite{bm2}. Furthermore, we will correct this
nuclear matter approximation by calculating the 2-particle 1-hole (2p1h)
term displayed in Fig.\ \ref{fig:diag}b
directly for the finite system, correcting the double-counting contained
in the Hartree--Fock term (see discussion below).

Using vector bracket transformation coefficients \cite{vecbr}, the $G$-matrix
elements obtained from the Bethe-Goldstone equation can be transformed from
the representation in coordinates of relative and center of mass momenta
to the coordinates of single-particle momenta in the laboratory frame in which
the 2-particle state would be described by quantum numbers such as
\begin{equation}
\left| k_1 l_1 j_1 k_2 l_2 j_2 J T \right\rangle ,  \label{eq:bra2p}
\end{equation}
where $k_i$, $l_i$ and $j_i$ refer to momentum and angular momenta of
particle $i$ whereas $J$ and $T$ define the total angular momentum and
isospin of the two-particle state. It should be noted that Eq. (\ref{eq:bra2p})
represents an antisymmetrized 2-particle state.
Performing an integration over one of
the $k_i$, one obtains a 2-particle state in a mixed representation of one
particle in a bound harmonic oscillator while the other is in a
plane wave state
\begin{equation}
\left| n_1 l_1 j_1 k_2 l_2 j_2 J T \right\rangle = \int_0^\infty dk_1 \, k_1^2
R_{n_1, l_1}(\alpha k_1) \, \left| k_1 l_1 j_1 k_2 l_2 j_2 J T\right\rangle
..
\label{eq:branp}
\end{equation}
Here $R_{n_1, l_1}$ stands for the radial oscillator function and the
oscillator length $\alpha = 1.72 $ fm$^{-1}$ has been selected. This
choice for the oscillator length corresponds to an oscillator energy of
$\hbar \omega_{osc}$ = 14 MeV. Therefore the oscillator functions
are quite appropriate to describe the wave functions of the bound
single-particle states in $^{16}$O. Indeed, it turns out that the
single-particle wave functions determined in self-consistent BHF calculations
for $^{16}$O have a large overlap with these oscillator functions\cite{carlo}.
Using the nomenclature defined in
Eqs. (\ref{eq:bra2p}) - (\ref{eq:branp}) our Hartree-Fock approximation
for the self-energy is easily obtained in the momentum representation
\begin{equation}
\Sigma^{HF}_{l_1j_1} (k_1,k'_1) =
\frac{1}{2(2j_1+1)} \sum_{n_2 l_2 j_2 J T} (2J+1) (2T+1)
\left\langle k_1 l_1 j_1 n_2 l_2 j_2 J T \right| G \left|
k'_1 l_1 j_1 n_2 l_2 j_2 J T\right\rangle .
\label{eq:selhf}
\end{equation}
The summation over the oscillator quantum numbers is restricted to the
states occupied in the IPM of $^{16}$O. This
Hartree-Fock part of the self-energy is real and does not depend on the
energy. It should be noted that the approximation, which is called
Hartree-Fock in this paper is not a Hartree-Fock approach in terms of the
bare NN interaction $V$. Using a realistic NN interaction $V$, as it is
done here, such a Hartree-Fock approach would yield a very unreasonable
description, in which the nucleons are typically unbound\cite{prog2}. What is
called Hartree-Fock approximation here, is defined in terms of a nuclear
matter $G$ matrix, which accounts already for NN correlations.

In order to evaluate now the single-particle Green function and densities
we consider a
 complete basis within a spherical box of a radius $R_{\rm box}$. This box
radius should be larger than the radius of the nucleus considered.
The calculated observables are independent of the choice of $R_{\rm box}$,
if it is chosen to be around 15 fm or larger. A complete and
orthonormal set of regular basis functions within this box is given by
\begin{equation}
\Phi_{iljm} ({\bf r}) = \left\langle {\bf r} \vert k_i l j m
\right\rangle = N_{il} j_l(k_ir)
{\cal Y}_{ljm} (\theta\phi ) \label{eq:boxbas}
\end{equation}
In this equation ${\cal Y}_{ljm}$ represent the spherical harmonics
including the spin degrees of freedom and $j_l$ denote the spherical
Bessel functions for the discrete momenta $k_i$ which fulfill
\begin{equation}
j_l (k_i R_{\rm box}) = 0 .
\label{eq:bound}
\end{equation}
Using the normalization constants
\begin{equation}
N_{il} =\cases{\frac{\sqrt{2}}{\sqrt{R_{\rm box}^3} j_{l-1}(k_i
R_{\rm box})}, & for $l > 0$ \cr
\frac{i \pi\sqrt{2}}{\sqrt{R_{\rm box}^3}}, & for $l=0$,\cr}
\label{eq:nbox}
\end{equation}
the basis functions defined in Eq. (\ref{eq:boxbas}) are orthogonal and
normalized within the box
\begin{equation}
\int_0^{R_{\rm box}} d^3 r\, \left\langle k_{i'} l' j' m' \vert {\bf r}
\right\rangle \left\langle {\bf r}
\vert k_i l j m \right\rangle
= \delta_{ii'} \delta_{ll'} \delta_{jj'} \delta_{mm'} .
\label{eq:boxi}
\end{equation}
Note that the basis functions defined for discrete values of the momentum
$k_i$ within the box differ from the plane wave states defined in the
continuum with the corresponding momentum just by the normalization constant,
which is $\sqrt{2/\pi}$ for the latter. This enables us to determine the
matrix elements of the nucleon self-energy in the basis of Eq.
(\ref{eq:boxbas})
from the results presented in Eq. (\ref{eq:selhf}).
For that purpose the Hartree-Fock Hamiltonian is diagonalized
\begin{equation}
\sum_{n=1}^{N_{\rm max}} \left\langle k_i \right| \frac{k_i^2}{2m}\delta_{in} +
\Sigma^{HF}_{lj} \left| k_n \right\rangle \left\langle k_n \vert a
\right\rangle_{lj} = \epsilon^{HF}_{a
lj} \left\langle k_i \vert a\right\rangle_{lj}. \label{eq:hfequ}
\end{equation}
Here and in the following the set of basis states in the box has been
truncated by assuming an appropriate $N_{\rm max}$. In the basis of
Hartree-Fock states $\left| a \right\rangle$, the Hartree-Fock propagator
is diagonal and given by
\begin{equation}
g_{lj}^{(0)} (a; E) = \frac{1}{E-\epsilon^{HF}_{a lj} \pm
i\eta} , \label{eq:green0}
\end{equation}
where the sign in front of the infinitesimal imaginary quantity $i\eta$ is
positive (negative) if $\epsilon^{HF}_{a lj}$ is above (below) the
Fermi energy. Knowing the expansion coefficients of the Hartree-Fock states
in the ``box-basis'' $\langle k_i \vert a\rangle_{lj}$ it is of course
trivial to present $g_{lj}^{(0)}$ in the basis of Eq. (\ref{eq:boxbas}).

The contributions to the self-energy, which are of second order in the
nuclear matter $G$ matrix are represented by the diagrams displayed in Figs.\
\ref{fig:diag}b) and \ref{fig:diag}c),
referring to intermediate 2p1h and 2-hole 1-particle (2h1p)
states, respectively. These two contributions define the correction to the
self-energy $\Delta\Sigma$ beyond the Hartree-Fock approximation in the
Dyson Eq. (\ref{eq:dyson}). These two contributions yield a complex and
energy-dependent correction to the self-energy.
The 2p1h contribution to the imaginary part is given by
\begin{eqnarray}
{W}^{2p1h}_{l_1j_1} (k_1,k'_1; E)
=& \frac{-1}{2(2j_1+1)}  \sum_{n_2 l_2 j_2} \sum_{l_3 l_4 j_3 j_4}
\sum_{J T} \int k_3^2 dk_3 \int k_4^2 dk_4  (2J+1) (2T+1) \nonumber \\
& \times \left\langle k_1 l_1 j_1 n_2 l_2 j_2 J T \right| G \left|
k_3 l_3 j_3 k_4 l_4 j_4 J T \right\rangle \nonumber \\
& \times  \left\langle k_3 l_3 j_3 k_4 l_4 j_4 T \right| G \left|
  k'_1 l_1 j_1 n_2 l_2 j_2 J T
  \right\rangle \nonumber \\
& \times \pi \delta\left(E +\epsilon_{n_2 l_2 j_2}
-\frac{k_3^2}{2m}-\frac{k_4^2}{2m}\right), \label{eq:w2p1h}
\end{eqnarray}
where the ``experimental'' single-particle energies $\epsilon_{n_2 l_2 j_2}$
are used for the hole states (-47 MeV, -21.8 MeV, -15.7 MeV for $s_{1/2}$,
$p_{3/2}$ and $p_{1/2}$ states, respectively), while the energies of the
particle states are given in terms of the kinetic energy only.
The expression in Eq. (\ref{eq:w2p1h}) still ignores the requirement that the
intermediate particle states must  be orthogonal to the hole states, which
are occupied for the nucleus under consideration. The techniques to incorporate
the orthogonalization of the intermediate plane wave states to the occupied
hole states as discussed in detail by Borromeo et al.\cite{boro} have also
been used here.
The 2h1p contribution to the imaginary part ${W}^{2h1p}_{l_1j_1}(k_1,k'_1; E)$
can be calculated in a similar way (see also \cite{boro}).

Our choice to assume pure kinetic energies for the particle states in
calculating the imaginary parts of $W^{2p1h}$ (Eq. (\ref{eq:w2p1h})) and
$W^{2h1p}$ may not be very realistic for the excitation modes at low energy.
Indeed a sizeable imaginary part in $W^{2h1p}$ is obtained only for
energies $E$ below -40 MeV. As we are mainly interested, however, in the
effects of short-range correlations, which lead to excitations of particle
states with high momentum, the choice seems to be appropriate.
A different approach would be required to treat the coupling
to the very low-lying two-particle-one-hole and two-hole-one-particle
states in an adequate way. Attempts at such a treatment can be found in
Refs.\ \cite{brand,rijsd,skou1,skou2}.

The real parts of the 2p1h and 2h1p terms in the self-energy can be
calculated from the corresponding imaginary parts by using dispersion
relations\cite{mahau}. As an example we present the dispersion relation
for the  2p1h part, which is given by
\begin{equation}
V^{2p1h}_{l_1j_1}(k_1,k_1';E)=\frac{P}{\pi} \int_{-\infty}^{\infty} \frac
{W^{2p1h}_{l_1j_1}(k_1,k_1';E')}{E'-E} dE', \label{eq:disper1}
\end{equation}
where $P$ means a principal value integral.
Since the Hartree--Fock contribution $\Sigma^{HF}$
has been calculated in terms of a
nuclear matter $G$-matrix, it already contains 2p1h terms of the kind displayed
in Fig.\ \ref{fig:diag}b).
Therefore one would run into problems of double counting if one
simply adds the real part $V^{2p1h}$ to the Hartree-Fock self-energy. Notice
that $\Sigma^{HF}$ does not contain any imaginary part because it is calculated
with a nuclear matter $G$-matrix at a starting energy for which $G$ is real.
In order to avoid such an over counting of the particle-particle
ladder terms, we subtract from the real part of the self-energy a
correction term $V_c$, which just contains this contribution calculated in
nuclear matter\cite{pap2}.

Summing up the various contributions we obtain the following expression for
the correction to the Hartree-Fock self-energy
\begin{equation}
\Delta\Sigma =  \left( V^{2p1h} - V_c + V^{2h1p}\right)
+ i\left( W^{2p1h} + W^{2h1p} \right) . \label{eq:defsel}
\end{equation}
Rather than calculating the various contributions to $\Delta\Sigma$ in the
basis of plane wave states, one may also calculate $\Delta\Sigma$ in the
box basis defined in Eq. (\ref{eq:boxbas}) or in the basis of Hartree-Fock
states $a,b$, which we have obtained from the diagonalization
in Eq. (\ref{eq:hfequ}). In this basis it is easy to determine the
so-called reducible self-energy, originating from an iteration of
$\Delta\Sigma$, by solving
\begin{equation}
\left\langle a \right| \Sigma^{red}_{lj}(E) \left| b \right\rangle =
\left\langle a \right| \Delta\Sigma_{lj}(E) \left| b \right\rangle
+ \sum_c
\left\langle a \right| \Delta\Sigma_{lj}(E) \left| c \right\rangle
g_{lj}^{(0)} (c ; E) \left\langle c \right| \Sigma^{red}_{lj}(E)
\left| b\right\rangle
\end{equation}
and obtain the propagator, which corresponds to the solution of the Dyson
Eq. (\ref{eq:dyson}) from
\begin{equation}
g_{lj} (a ,b ;E ) = g_{lj}^{(0)} (a;E )
\left\langle a \right| \Sigma^{red}_{lj}(E) \left| b \right\rangle
g_{lj}^{(0)} (b ;E ) .
\end{equation}
Using this representation of the Green function one can calculate the
continuum contribution to the density matrix in the ``box basis'' from
\begin{equation}
\rho_{lj}^c (k_1, k_2) = \frac{1}{\pi} \int_{-\infty}^{E_{2h1p}} dE\,
\mbox{Im} \left(
\sum_{a, b} \left\langle k_1 \vert a \right\rangle_{lj} g_{lj}
(a ,b ;E )
\left\langle b \vert k_2 \right\rangle_{lj}\right) , \label{eq:skob}
\end{equation}
where the integration limit $E_{2h1p}$ refers to the negative threshold of
2h1p states for the $lj$ combination considered. In our studies this energy
is well below the corresponding Hartree-Fock energies. Therefore the
imaginary part of $g_{lj}$ in the region of integration is different from
zero only due to the imaginary part in $\Sigma^{red}$, which just describes
the coupling of the quasi particle states to the 2h1p continuum.

Besides this continuum contribution, the hole spectral function
also receives contributions from the quasihole states \cite{wimref}. The
energies and wave functions of these quasihole states can be determined
by diagonalizing the Hartree-Fock Hamiltonian plus $\Delta\Sigma$ in the
``box basis''
\begin{equation}
\sum_{n=1}^{N_{\rm max}} \left\langle k_i \right| \frac{k_i^2}{2m}\delta_{in} +
\Sigma^{HF}_{lj} + \Delta\Sigma_{lj} (E=\epsilon^{qh}_{\Upsilon lj})
\left| k_n \right\rangle \left\langle k_n \vert \Upsilon \right\rangle_{lj} =
\epsilon^{qh}_{\Upsilon lj} \left\langle k_i \vert \Upsilon \right\rangle_{lj}
. \label{eq:qhequ}
\end{equation}
Since in the present work $\Delta\Sigma$ only contains a sizeable imaginary
part for energies $E$ below $\epsilon^{qh}_{\Upsilon}$, the
energies of the quasihole states come out real and the continuum contribution
is separated in energy from the quasihole
contribution. The quasihole contribution to the density matrix is
given by
\begin{equation}
\rho^{qh}_{\Upsilon lj} (k_1, k_2) = Z_{\Upsilon lj}
{\left\langle k_1 \vert \Upsilon \right\rangle_{lj} }
{\left\langle \Upsilon \vert k_2 \right\rangle_{lj} }
, \label{eq:skoqh}
\end{equation}
with the spectroscopic factor for the quasihole state given by \cite{wimref}
\begin{equation}
Z_{\Upsilon lj} =
\bigg( {1-{\partial \left\langle \Upsilon \right| \Delta\Sigma_{lj}(E)
\left| \Upsilon \right\rangle \over
\partial E} \bigg|_{\epsilon^{qh}_{\Upsilon lj}}} \bigg)^{-1} .
\label{eq:qhs}
\end{equation}
Finally, the total single-particle density matrix is given as the sum of
the continuum part of Eq. (\ref{eq:skob}) and the quasihole contributions
of Eq. (\ref{eq:skoqh}) summed over all quasihole states with an energy
below the Fermi energy $\epsilon_F$
\begin{equation}
\rho_{lj} (k_1, k_2) = \rho_{lj}^c (k_1, k_2) + \sum_{
\epsilon^{qh}_{\Upsilon lj}<\epsilon_F}
\rho^{qh}_{\Upsilon lj} (k_1, k_2)\ .     \label{eq:rhobox}
\end{equation}
At this stage it is now of course straightforward to determine the
natural orbits and their occupation numbers by diagonalizing this density
matrix according to Eq. (\ref{eq:natorb}).

\section{Results and Discussion}

\subsection{Natural orbits and occupations}
The numerical results to be discussed in this section have been obtained
for each combination of orbital angular momentum $l$ and total angular
momentum $j$ by using a
basis of single-particle states defined according to Eqs.
(\ref{eq:boxbas}) - (\ref{eq:nbox}) for a spherical box with a radius $R$
of 20 fm. Basis states up to $N_{\rm max}$=20 were taken into account. With
these assumptions the density matrix for each single-particle channel $lj$
is represented by a matrix of dimension $N_{\rm max}$=20. The diagonalization
of these matrices yields eigenvalues which are the occupation probabilities
$n_{lj\alpha}$ of the natural orbits (see Eq.(\ref{eq:natorb})) and the
eigenvectors correspond to the expansion coefficients of these natural
orbits ($\phi_{lj\alpha}$) in the chosen basis.

Results for these occupation probabilities are displayed in table
\ref{tab:nnor} ordered for each partial wave ($lj$) according to the
magnitude of the occupation probability. Here and in the following we will
use the notation that the natural orbit identified by $\alpha$=1 corresponds
to the orbit with largest occupation probability and also the orbits with
$\alpha$=2...20 are ordered with respect to the occupation probability
$n_\alpha$. Note that natural orbits with $\alpha$ larger than 4 show an
occupation which is so close to zero, that their contribution to the
total occupations for each $lj$, which are also given in table \ref{tab:nnor},
is hardly visible within the accuracy used in that table.

For those partial waves $lj$, for which one orbit is
completely occupied in the Hartree-Fock (HF) approach ($s_{1/2}$, $p_{3/2}$,
and $p_{1/2}$ in our example $^{16}$O), one finds that the first natural
orbit ($\alpha$=1) exhibits an occupation probability very close to one,
while the occupation probabilities of the natural orbits for $\alpha$ =2
and larger are smaller by about a factor of 0.01.
Similar observations have been reported for the natural orbits of $^{3}$He
drops containing 70 atoms.\cite{lewar}
It is interesting to observe that the eigenvalues for the `occupied'
orbits in \cite{lewar} are in general 10-40\% smaller than the corresponding
ones calculated here for $^{16}$O, indicating that correlations
between $^{3}$He atoms are stronger than those between nucleons.
For those orbits $lj$ which are
completely unoccupied in the HF approach, like $d_{5/2}$, $d_{3/2}$ or
$f_{7/2}$ in our example, the occupation probabilities are of course much
smaller
in agreement with the results in \cite{lewar}.
The ratio $n_{\alpha=2}$ versus $n_{\alpha=1}$ for these orbits
is not as small as in the case of the orbits with occupation in the HF
approach. As we will discussed more in detail below, this difference is due
to the quasihole contribution to the density in these ``occupied'' orbits,
which is missing in the $lj$ channels which don't show any occupation in HF.

The total occupation of these orbits with $l>1$ is decreasing with
increasing $l$. This could
be interpreted to reflect the feature that orbits which have a larger
excitation energy within the shell-model are less occupied. It is
remarkable, however, that for a given $l$ the states with $j=l-1/2$ are
more populated than those with $j=l+1/2$, although the spin-orbit term of the
shell-model yields less binding for the former than for the latter states.
This is true for the example of the $d$ states displayed in the table but
also for states with $l>2$, which are not given explicitly here. In order
to understand this feature we must recall that the occupation probability
for e.g. natural orbits with $d_{5/2}$ symmetry in  $^{16}$O, reflects the
probability to generate in knock-out reactions states of the $A-1$ particle
system
\begin{equation}
|\Psi^{(A-1)}_{lj} >\,\leftarrow\, a_{lj}|\Psi^{(A)}_0> \ . \label{eq:psia-1}
\end{equation}
The resulting state $|\Psi^{(A-1)}_{lj} >$ is a one-hole state with respect
to the correlated ground state of $^{16}$O, it does not contain, however,
any component with one hole in the $d_{5/2}$ orbit with respect to the
HF ground state configuration for $^{16}$O.
Such configurations do not exist or do not have the
particle number $(A-1)$=15. The occupation probability of natural orbits
with $d_{5/2}$ symmetry reflects the possibility of generating states with
two-hole 1-particle or more complicate (n+1)-hole n-particle configurations
as compared to the HF ground state configuration for $^{16}$O. This
probability, however, is not affected by single-particle energies of the
$d_{5/2}$ orbits but rather determined by the number and energies of these
(n+1)-hole n-particle configurations and the probability of these
configurations to be generated by applying $a_{lj}$ to the correlated
ground state $\Psi^{(A)}_0$. Keeping this in mind the occupation
probabilities of the natural orbits with $l>1$ are easily understood.

The radial shapes of natural orbits with $\alpha$=1 to 3 are displayed in
Fig. \ref{fig:natr}. At first sight these natural orbits look like solutions
of a Schr\"odinger equation for a particle moving in a single-particle
potential. They exhibit the same $l$-dependence ($\phi_{lj\alpha}\sim r^l$)
for small values of $r$, the number of nodes is essentially identical to
$\alpha -1$, and their values decrease for large distances as it is the case
for single-particle wave functions for bound states.

In order to characterize these natural orbits with different $\alpha$ on a
more quantitative level, we have calculated the square radii of these orbits
\begin{equation}
<r^2>_{lj\alpha} = \int r^2 dr\,\phi_{lj\alpha}^2(r)\, r^2
\end{equation}
and show some typical results in table \ref{tab:rad}. It should be emphasized
that the resulting $<r^2>_{lj\alpha}$ do not increase with the number of
nodes, i.e.\ the value of $\alpha$, as one expects from single-particle wave
functions in a potential. Assuming e.g.\ an oscillator potential the
corresponding values for $<r^2>$ increase as a function of $\alpha$
proportional to $(2\alpha + l -1/2)$. The results for the
natural orbits are much less increasing with $\alpha$, in the case of the
$l=1$ orbits the values obtained for $\alpha=2$ are even smaller than the
corresponding results for $\alpha=1$. This means that the natural orbits
with $\alpha >1$ yield a density distribution much more localized in the
center of the nucleus than would be obtained  from corresponding
single-particle wave functions for a local potential.

Furthermore it is
remarkable that the values of $<r^2>$ obtained for the $\alpha=1$ natural
orbits with angular momenta $l=2$ and $l=3$ are smaller then the radii
calculated for the natural orbits with large occupation probability and
$l=1$. This difference is due to the fact that the $l=1,\ \alpha=1$
natural orbits are dominated by the quasihole contribution to the density
matrix (see Eq.\ \ref{eq:skoqh}). The quasihole wave functions have overlaps
larger than 0.99 with the corresponding natural orbits $\alpha=1$. The
quasihole states contribute an occupation of 0.78, 0.914 and 0.898 to the
total occupations of 0.921, 0.947 and 0.93 for the $\alpha=1$ natural
orbits with $l_j$ equal to $s_{1/2}$, $p_{3/2}$ and $p_{1/2}$, respectively.
The radii of the quasihole states are slightly larger than the radii of the
corresponding natural orbits (see table \ref{tab:rad} and also figure
\ref{fig:qhfno}). This
indicates that the continuum contribution to the natural orbits tend to
have smaller radii than the quasihole part. Therefore it can also be
understood that radii of natural orbits with $l>1$, which do not
contain any quasihole contributions, can be smaller than those for $l=1$.

 From a more general point of view, however, one may say that for all
those orbits, which are occupied within the naive shell-model and
therefore yield a quasihole state, the radial wave function for the
natural orbit with largest occupation probability ($\alpha =1$) is very
similar to the wave function of the quasihole state. Furthermore, both wave
functions are very close to the corresponding single-particle wave
functions obtained in the HF approach (see figure \ref{fig:qhfno}).

Summarizing this part of the discussion we find that for a given $lj$ the
continuum contribution to the density matrix (see Eq. \ref{eq:skob}) yields
smaller radii than the corresponding quasihole wave function. Even if we
sum the contributions of all natural orbits
\begin{equation}
<r^2>_{lj} = \frac{\sum_\alpha n_{lj\alpha}<r^2>_{lj\alpha}}{
\sum_\alpha n_{lj\alpha}}
\end{equation}
the result is smaller than just the radius of the quasihole state (see table
\ref{tab:rad}). In order
to obtain the total radius of the nucleus one has to sum over all partial
waves $lj$, multiply each $<r^2>_{lj}$ by the total occupation of these
orbits $\sum_\alpha n_{lj\alpha}$ times the degeneracy ($2(2j+1)$) and
divide by the total particle number. Due to the depletion of states with
$l\leq 1$ accompanied by a small occupation of orbits with $l>1$ one
obtains a value for the total radius (R=2.55 fm), which is only slightly
below the result obtained in the Hartree-Fock approach (R=2.59 fm).

These features are also displayed in figure \ref{fig:todens}, where the
local density distribution obtained in HF (long dashes) is compared to the
total density distribution (solid line). Also we present the contribution
to the total density distribution coming from partial waves with $l> 1$
(short dashes). One finds that the depletion of states with small $l$
(in particular $l=0$) yields a reduction of the density in the center of
the nucleus. The difference between the density distribution from the
quasi-hole approximation (dashed-dotted line) and the total result is
due to the continuum contribution to the density.

\subsection{Spectral decomposition}
In the preceding discussion it has been pointed out already that the
natural orbits with occupation probabilities close to one are dominated
by the quasihole contribution. This means that features of these natural
orbits are explored in knock-out experiments like $(e,e'p)$ populating
the states of low energy in the daughter nucleus. In the following we
would like to discuss the question, to which extent the analysis of
experiments with large missing energies, leading to high excitation
energies in the residual $(A-1)$ system, exhibits the properties of the
natural orbits with smaller occupation probabilities.

\begin{figure}
\caption{Natural orbits resulting from the diagonalization of the
continuum contribution to the single-particle density in $l=0$ partial
wave. The natural orbits obtained from the energy interval $[-100
MeV,\epsilon_{F}]$ in Eq.(19) (short dashes) and those from the interval
$[-\infty ,-100 MeV]$ (long dashes) are compared to those from the
whole energy range (solid line). The left part of the figure shows the
natural orbits with $\alpha =1$, while the right part displays those
for $\alpha =3$.}
\label{fig:conti}
\end{figure}
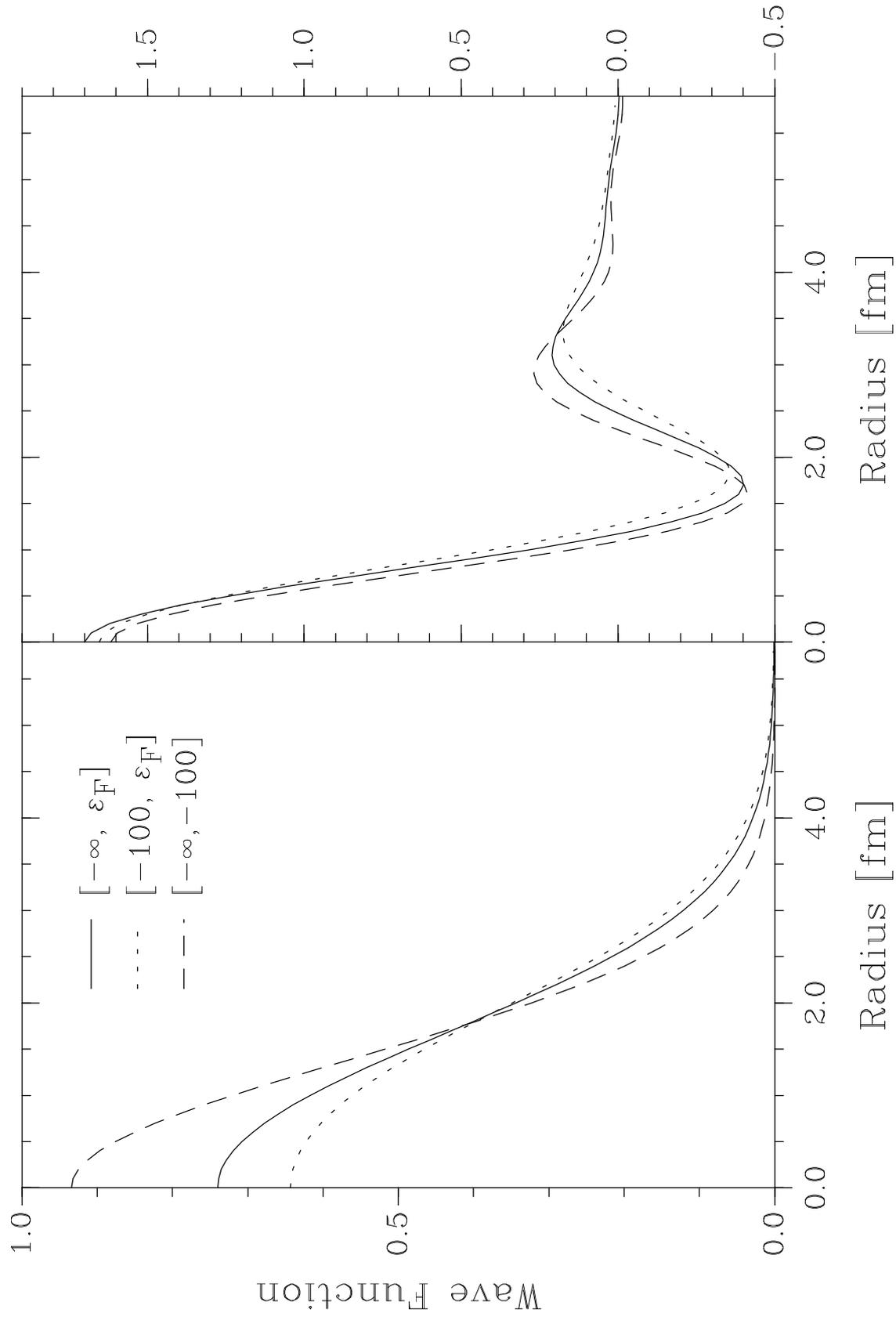

For that purpose we analyze the continuum contribution to the density
matrix ($\rho^c$, see eq.(\ref{eq:skob})). For the example of orbits
with $l=0$, the largest occupation numbers resulting from the diagonalization
of this continuum part of the density matrix are listed in table
\ref{tab:conti} and the corresponding natural orbits are displayed in
figure \ref{fig:conti}. The natural orbit with largest occupation,
obtained from this continuum part of the density matrix, yields a
radial  wave function, which is again very similar to the corresponding
quasihole state. In fact, the overlap of the natural orbit displayed in
the left part of figure \ref{fig:conti} has an overlap with the
quasihole wave function of the $l=0$ state, which is as large as 0.997.
This means that also the continuum part of the density matrix is
characterized by a dominant natural orbit very similar to the quasihole
state.

In a next step we split the energy integration of eq.(\ref{eq:skob})
into 2 parts. The first interval ranges from $E =-100$ MeV to the
threshold of the $2h1p$ continuum, while the second interval covers
energies from $-\infty$ to $E =-100$ MeV. Restricting the energy
integration to one of these two intervals we can study either the
single-particle density to be observed in knock-out experiments with
very large missing energies (interval 2) or at medium missing energies.
Eigenvalues (i.e.~occupation probabilities $n_{\alpha}$) and
eigenvectors (natural orbits) of these parts of the density matrix are
also presented in table \ref{tab:conti} and figure \ref{fig:conti},
respectively.

The natural orbits originating from the energy interval 2 (large
missing energies) are more localized in the center of the nucleus than
those derived from the energy interval which corresponds to lower
excitation energies in the residual system. The overlaps of the various
natural orbits with the same $\alpha$, however, are still large
(typically larger than 0.95). What seems to be more significant is the
relative importance of the different energy intervals for the various
natural orbits. The occupation of the $\alpha=1$ natural orbit mainly
originates from the energy interval 1, whereas the natural orbits with
$\alpha$ equal to 2 or larger, collect the main part of the occupation
from energies below -100 MeV in the integral of eq.(\ref{eq:skob}).

The same features are also observed in the other partial waves as can be
seen from the results displayed in table \ref{tab:exspec}. In this
table we give for various partial waves $lj$ the occupation
probabilities of the natural orbits derived from the total density
matrix, split into the contribution resulting from the quasihole
part of the density matrix (denoted by $n^{qh}_{\alpha}$) and the continuum
part ($n^c_{\alpha}$). These results show again that the $\alpha =1$
natural orbits are dominated by the quasihole component of the density
matrix, while the other natural orbits mainly originate from the
continuum contribution to the single-particle density.

Furthermore we have listed for each natural
orbit, $\phi_{lj\alpha}$, the mean value for the energy in the 2h1p
continuum (see also Eq.(\ref{eq:skob}))
\begin{equation}
\omega_{lj\alpha}=\frac{1}{\pi n^c_{\alpha}}
\int_{-\infty}^{E_{2h1p}} dE\, E \, \mbox{Im} \left(
\sum_{k_{1},k_{2}} \left\langle \alpha\vert k_1 \right\rangle_{lj} g_{lj}
(k_{1} ,k_{2} ;E )
\left\langle k_2 \vert \alpha \right\rangle_{lj}\right) .
\label{eq:omeg}
\end{equation}
In this equation $g_{lj}(k_{1} ,k_{2} ;E )$ stands for the
single-particle Green function in the ``box-basis'' of Eqs.(\ref{eq:boxbas}) -
(\ref{eq:nbox}) and $\langle k_i \vert \alpha \rangle_{lj}$ denote the
expansion coefficients of the natural orbits in this basis.

Inspecting these mean values for the energies one finds that the
absolute values of these energies increase with $\alpha$, the index of
the natural orbits chosen with respect to the occupation probability.
This implies that the natural orbits, which have radial wave functions
with a larger number of nodes ($\alpha >1$, see figure \ref{fig:natr}),
``collect'' the occupation at more negative energies than the natural
orbits with $\alpha =1$.

Also we list in table \ref{tab:exspec} the continuum contribution to
the expectation values for the kinetic energies of the natural orbits
\begin{equation}
t_{lj\alpha}=\frac{1}{\pi n^c_{\alpha}} \int_{-\infty}^{E_{2h1p}} dE\,
\left\langle k_{1}\vert T_{kin}\vert k_2 \right\rangle_{lj} \,
\mbox{Im} \left(
\sum_{k_{1},k_{2}} \left\langle \alpha\vert k_1 \right\rangle_{lj} g_{lj}
(k_{1} ,k_{2} ;E )
\left\langle k_2 \vert \alpha \right\rangle_{lj}\right) .
\label{eq:tlj}
\end{equation}
Also the expectation values $t_{lj\alpha}$ increase with $\alpha$. For
$\alpha$=1 they are slightly larger than the corresponding expectation
values for the quasihole states. This is in line with the observation
made above that the continuum part of the density matrix tends to
decrease the radius of the natural orbit with $\alpha=1$ as compared to
the quasihole state. The expectation values for $\alpha$ larger than
one increase very drastically with $\alpha$ since the corresponding natural
orbits contain strong components with high momenta as can bee seen from
the radial shapes in figure \ref{fig:natr}.

Concluding this part of the discussion we find that the natural orbits
with $\alpha$ larger than 1 (indicating their small occupations)
represent the high momentum components in the single-particle density
and originate mainly from the imaginary part of the Green function at
very negative energies. This means that the features of these natural
orbits might be explored by analyzing knock-out experiments with large
missing energies.

The occupation probabilities and expectation values listed in table
\ref{tab:exspec} can also be used to calculate the total energy of the
nucleus $^{16}$O via the ``Koltun sum rule''
\begin{equation}
{\cal E} = \sum_{lj} 2 *(2j+1) \left[ \sum_{\Upsilon} \frac{Z_{\Upsilon
lj}}{2} \left( \epsilon^{qh}_{\Upsilon lj} + t^{qh}_{\Upsilon lj}
\right ) \, + \, \sum_{\alpha} \frac{n^c_{\alpha}}{2} \left(
\omega_{lj\alpha} + t_{lj\alpha} \right) \right] \label{eq:koltun}
\end{equation}
The quasihole contribution to the total energy is calculated in terms
of the quasihole energy $\epsilon^{qh}_{\Upsilon lj}$ (see
Eq.(\ref{eq:qhequ})), the kinetic energy for the quasihole states
$t^{qh}_{\Upsilon lj}$ and the spectroscopic factors $Z_{\Upsilon lj}$
(see Eq.(\ref{eq:qhs})). It is remarkable that this quasihole
contribution yields only around 37 percent of the total energy although
the quasihole part of the single-particle density is responsible for 89
percent of the total particle number. The largest contribution to the
energy in Eq.(\ref{eq:koltun}) originates from the continuum part of
the natural orbits with $\alpha=1$. These states contribute more than
58 percent of the total energy although the particle number of these
contributions is only around 9 percent. The remaining part of the total
energy ${\cal E}$, which is -3 MeV out of the total calculated ${\cal
E}=-81.1 MeV$ (see also \cite{pap2}), is due to the natural orbits with
$\alpha$ larger than 1.

It is worth repeating that a large part of the density matrix can be
described in terms of one natural orbit for each partial wave. These
contributions of the $\alpha=1$ natural orbits account for 97.9 percent
of the total particle number and for 96 percent of the total energy.
Therefore the description of the
single-particle density matrix in terms of these natural orbits with
largest occupation probabilities should be sufficient for the
calculation of many single-particle observables. It should be kept in
mind that already this description in terms of one natural orbit in
each partial wave is beyond the mean-field approximation: (i) It allows
for a depletion of orbits, which are completely occupied in the
mean-field approach, and occupation of partial waves, which are not
occupied in the mean-field picture. (ii) This approach takes into
account the most important part of the spectral distribution of the
single-particle density matrix (see the contributions to the energy in
Eq.(\ref{eq:koltun})). The restriction to the natural orbits with
$\alpha$=1, however, should not be sufficient to obtain a reliable
description of the high-momentum components in the single-particle
density matrix. These components are described mainly in terms of the
natural orbits with smaller occupation probability.

\section{Conclusions}
Properties of the
correlated wave function for the nucleus $^{16}$O, which includes
the effects of short-range correlations derived from a realistic NN
interaction by means of the Green function method\cite{pap2}, have been
analyzed in terms of natural orbits. It turns out that the
single-particle density matrix can very well be described in terms of a
few natural orbits for each partial wave $lj$. If one restricts the
representation of the density matrix in Eq.(\ref{eq:natorb}) to only one
orbit per partial wave (the orbit with largest occupation probability
$n_{\alpha}$, here referred to as the $\alpha=1$ natural orbits), one
obtains a density matrix, which accounts already for 97.9 percent of
the total particle number. These $\alpha=1$ natural orbits have a large
overlap with the wave functions of the quasihole states as well as the
Hartree-Fock (HF) single-particle wave functions for those orbits, which are
occupied within the mean-field approximation.

At first sight also the natural orbits in partial waves, which are
unoccupied in the HF approximation, and the natural orbits with small
occupation probabilities ($\alpha >1$) look like single-particle wave
functions for a nucleons in an excited state of a single-particle
potential. A closer inspection shows, however, that the radii for these
``excited'' orbits are much smaller than those obtained for a motion
within a single-particle orbit. These $\alpha >1$ natural orbits mainly
originate from the single-particle Green function at large negative
energies ($E<$-100 MeV). Although the occupation of these orbits is
much smaller than those with $\alpha$=1, these orbits describe the
high-momentum components of the single-particle density matrix.
Therefore in order to study these ``excited'' natural orbits or the
high-momentum components of the single-particle density, one has to
analyze nucleon knock-out experiments with large missing energy.

For the understanding of the bulk properties of nuclei, like binding
energy and radius, it seems to be a good approximation to restrict the
study to natural orbits with maximal occupation probability. It is not
sufficient, however, to restrict the evaluation of the energy according
to the ``Koltun sum rule'' (Eq.\ref{eq:koltun}) to the contributions of
the quasihole states. The main contributions arise from the
single-particle Green function at energies deep in the 2h1p continuum.
In order to improve the present studies towards a self-consistent
treatment of the single-particle Green function, it should be a good
approximation to describe this Green function in terms of the natural
orbits with largest occupation probabilities.

This research project has partially been supported by SFB 382 of the
"Deutsche Forschungsgemeinschaft", DGICYT, PB92/0761
(Spain), the EC-contract CHRX-CT93-0323, and the U.S. NSF under Grant No.
PHY-9307484. One of us (H.M.) is pleased
to acknowledge the warm hospitality at the Facultat de Fisica,
Universitat de
Barcelona, and the support by the program for Visiting Professors of this
University.

\begin{table}[h]
\caption{Occupation of natural orbits in $^{16}$O.
Listed are the occupation probabilities $n_{\alpha}$ (see eq.1) of those
4 natural orbits, which show the largest occupation probabilities in various
partial waves. Also given are the total occupation probabilities, i.e. the
sum over all $n_{\alpha}$ including orbits with $\alpha > 4$.}
\label{tab:nnor}
\begin{center}
\begin{tabular}{c|rrrrrr}
&&&&&&\\
${\alpha}$&\multicolumn{1}{c}{$s_{1/2}$}&\multicolumn{1}{c}{$p_{3/2}
$}&\multicolumn{1}{c}{$p_{1/2}$}&
\multicolumn{1}{c}{$d_{5/2}$}&\multicolumn{1}{c}{$d_{3/2}$}&
\multicolumn{1}{c}{$f_{7/2}$}
 \\
&&&&&&\\ \hline
&&&&&&\\
1 & 0.9206 & 0.9467 & 0.9304 & 0.0154 & 0.0189 & 0.0056 \\
2 & 0.0127 & 0.0073 & 0.0083 & 0.0033 & 0.0045 & 0.0013 \\
3 & 0.0024 & 0.0014 & 0.0019 & 0.0007 & 0.0010 & 0.0003 \\
4 & 0.0004 & 0.0003 & 0.0004 & 0.0001 & 0.0001 & 0.0000 \\
&&&&&&\\
$\sum_\alpha n_{\alpha}$ & 0.936 & 0.956 & 0.941 & 0.0195 & 00245 & 0.0073 \\
&&&&&&\\
\end{tabular}
\end{center}
\end{table}
\begin{table}[h]
\caption{Radii of Natural Orbits. Listed are the expectation values of
$r^2$ for natural orbits with largest occupation probabilities in various
partial waves. Also given are the expectation value for the total
single-particle density ($<r^2>_{lj}$, see Eq. (25)) in each partial
wave (normalized to 1) and the
corresponding results for the Hartree-Fock (HF) and the quasihole states
(qh). All entities listed are given in fm$^2$.}
\label{tab:rad}
\begin{center}
\begin{tabular}{c|rrrrrr}
&&&&&&\\
${\alpha}$&\multicolumn{1}{c}{$s_{1/2}$}&\multicolumn{1}{c}{$p_{3/2}
$}&\multicolumn{1}{c}{$p_{1/2}$}&
\multicolumn{1}{c}{$d_{5/2}$}&\multicolumn{1}{c}{$d_{3/2}$}&
\multicolumn{1}{c}{$f_{7/2}$}
 \\
&&&&&&\\ \hline
&&&&&&\\
1 & 4.092 & 7.159 & 7.535 & 5.746 & 5.967 & 6.270 \\
2 & 4.446 & 5.761 & 6.070 & 7.033 & 7.215 & 9.070 \\
3 & 6.978 & 9.981 & 10.364 & 14.105 & 13.563 & 16.296 \\
4 & 17.982 & 20.445 & 20.172 & 22.776 & 22.872 & 26.877 \\
&&&&&&\\
$<r^2>_{lj}$ & 4.112 & 7.155 & 7.534 & 6.370 & 6.582 & 7.392 \\
HF & 3.974 & 7.435 & 7.963 &&&\\
qh & 4.164 & 7.258 & 7.653 &&&\\
&&&&&&\\
\end{tabular}
\end{center}
\end{table}
\begin{table}[h]
\caption{Occupation probabilities for natural orbits obtained by
diagonalizing the continuum part of the density matrix for the $l=0$
partial wave. Results are
presented restricting the energy integral of Eq.(19) to the intervals
shown in the first line.}
\label{tab:conti}
\begin{center}
\begin{tabular}{c|rrr}
&&&\\
${\alpha}$&\multicolumn{1}{c}{$[-100 \mbox{MeV},\epsilon_{F}]$}
&\multicolumn{1}{c}{$[-\infty ,-100 \mbox{MeV}]$}
&\multicolumn{1}{c}{$[-\infty ,\epsilon_{F}]$}\\
&&&\\ \hline
&&&\\
1 & 0.09390 & 0.04920 & 0.14118 \\
2 & 0.00465 & 0.00627 & 0.01232 \\
3 & 0.00058 & 0.00121 & 0.00222 \\
4 & 0.00006 & 0.00018 & 0.00037 \\
&&&\\
$\sum_{\alpha} n_{\alpha}$ & 0.09919 & 0.05693 & 0.15612 \\
&&&\\
\end{tabular}
\end{center}
\end{table}
\begin{table}[h]
\caption{Expectation of natural orbits in various partial waves
$l_{j}$. Listed are the part of the occupation probabilities arising
from the quasihole ($n^{qh}_{\alpha}$) and the continuum
($n^{c}_{\alpha}$) contribution to the single-particle density,
the mean value for the energy ($\omega_{lj\alpha}$, see Eq.(27)) and the
kinetic energy ($t_{lj\alpha}$, Eq.(28)) in the 2h1p continuum.
Furthermore we also give the single-particle energy ($\epsilon^{qh}$)
and the kinetic energy $t^{qh}$ of the quasihole states, if they exist.}
\label{tab:exspec}
\begin{center}
\begin{tabular}{c|rrrr}
&&&&\\
$\alpha$&\multicolumn{1}{c}{$n^{qh}_{\alpha}$} &
\multicolumn{1}{c}{$n^{c}_{\alpha}$} &
\multicolumn{1}{c}{$\omega_{lj\alpha}$ [MeV]} &
\multicolumn{1}{c}{$t_{lj\alpha}$ [MeV]}\\
&&&&\\ \hline
&&&&\\
&\multicolumn{4}{c}{$s_{1/2}$\ \ \ \ $\epsilon^{qh}$ = -34.30 MeV, \ \
\ \ $t^{qh}$ = 11.23 MeV}\\
&&&&\\
1 & 0.77988 & 0.14068 & -87.13 & 12.49 \\
2 & 0.00006 & 0.01268 & -118.22 & 50.28 \\
3 & 0.00002 & 0.00234 & -126.71 & 95.49 \\
&&&&\\
&\multicolumn{4}{c}{$p_{3/2}$\ \ \ \ $\epsilon^{qh}$ = -17.90 MeV, \ \
\ \ $t^{qh}$ = 18.06 MeV}\\
&&&&\\
1 & 0.91420 & 0.03251 & -87.22 & 23.70 \\
2 & 0.00009 & 0.00718 & -122.75 & 64.32 \\
3 & 0.00001 & 0.00139 & -126.26 & 98.74 \\
&&&&\\
&\multicolumn{4}{c}{$d_{5/2}$}\\
&&&&\\
1 &  & 0.01539 & -95.65 & 44.44 \\
2 &  & 0.00332 & -115.89 & 89.77 \\
3 &  & 0.00073 & -128.97 & 143.07 \\
&&&&\\
\end{tabular}
\end{center}
\end{table}
\clearpage
\begin{figure}
\caption{Graphical representation of the Hartree-Fock (a), the 2-particle
1-hole (2p1h, b) and the 2-hole 1-particle contribution (2h1p, c) to the
self-energy of the nucleon}
\label{fig:diag}
\end{figure}
\begin{figure}
\caption{The radial shape of natural orbits with $l_j$ = $s_{1/2}$, $p_{3/2}$
and $d_{5/2}$. Ordered with respect to the occupation number, results are
displayed for $\alpha$ = 1 (solid line), 2 (long dashes) and 3 (short
dashes)}
\label{fig:natr}
\end{figure}
\begin{figure}
\caption{Orbital shape of the natural orbit ($\alpha =1$, solid line) the
Hartree-Fock (short dashes) and the quasihole wave function in the $s_{1/2}$
partial wave}
\label{fig:qhfno}
\end{figure}
\begin{figure}
\caption{Density distribution of nucleons in $^{16}$O as a function of
the distance $r$ from the center of the nucleus. The result obtained in
the HF approximation (long dashes) is compared to the one obtained
within the Green function approach (solid line). Also displayed are the
contributions of the quasi-hole states to this total density
(dash-dotted line) and the density originating from the occupation of
states with $l$ larger than 1 (short dashes).}
\label{fig:todens}
\end{figure}
\end{document}